\documentclass[a4paper]{article}
\usepackage{amsmath}
\usepackage{amssymb}
\usepackage{esint}

\newtheorem{theorem}{Theorem}[section]
\newtheorem{lemma}{Lemma}[section]
\newtheorem{proposition}{Proposition}[section]

\def\tire{\thinspace--\thinspace} 

\title{Restricted three particle quantum walk on ${\bf Z_{\bf +}}$: explicit
solution}

\author{R.~Iasnogorodski\thanks{State Chemical-Pharmaceutical University, Prof. Popov str., 14, Saint Petersburg, 197376, Russia, E-mail: iasnogorodski@mail.ru}
\and
V.~Malyshev\thanks{Mechanics and Mathematics Faculty, Lomonosov Moscow State University, Leninskie
Gory~1, Moscow, 119991, Russia, E-mail: 2malyshev@mail.ru}
\and
A.~Zamyatin\thanks{Mechanics and Mathematics Faculty, Lomonosov Moscow State University, Leninskie
Gory~1, Moscow, 119991, Russia, E-mail: andrei.zamyatin@mail.ru}
}

\date{}

\begin{document}

\maketitle

\begin{abstract}
We consider 3 particles on ${\bf Z}{}_{+}$. One of them has infinite
mass and stands still at $0$. The particles interact only if all
of them are at the point $0$. We give a full description of essential,
point and discrete spectra of the corresponding hamiltonian. 
\end{abstract}

\section{Introduction }

The development of the theory of two-dimensional Wiener\tire Hopf equations goes back  to 1969 (see \cite{mal_1969}). It was a mixture of various
methods: functional equations on algebraic curves, elliptic curves
and Galois theory, integral equations. In the review \cite{mal_1976}
it was explained why such methods are crucial for one-dimensional 3-particle random walk problem.

In this paper we present the first application of these methods to the theory of quantum
walks.

\section{Model and main result}

\paragraph*{Hamiltonian}

We consider 3 particles on ${\bf Z}_{+}$. One of them stands still
at $0$ (has infinite mass) and the state of other 2 particles is
defined by complex wave function $f=\{f_{m,n}\},m,n\in{\bf Z}_{+},\,f_{m,n}\in{\bf C},$
from $l_{2}({\bf Z}_{+})\otimes l_{2}({\bf Z}_{+})=l_{2}({\bf Z}_{+}^{2})$,
that is we assume that 
\[
\sum_{n,m=0}^{\infty}|f_{n,m}|^{2}<\infty
\]
We can represent $f$ differently 
\[
f=\sum_{n,m=1}^{\infty}f_{m,n}e_{m,n}
\]
where $e_{m,n},(m,n)\in{\bf Z}_{+}^{2},$ is the standard orthonormal
basis in $l_{2}({\bf Z}_{+}^{2})$.

The evolution is defined by the unitary group $e^{itH}$ with the
Hamiltonian $H=H_{0}+V$, where $H_{0}$ and $V$ are the following
linear bounded selfadjoint operators:
\[
H_{0}e{}_{m,n}=-\lambda(e_{m+1,n}+e_{m-1,n}+e_{m,n+1}+e_{m,n-1}),\:n,m\geq1
\]
\[
H_{0}e{}_{0,n}=-\lambda(e_{1,n}+e_{0,n+1}+e_{0,n-1}),\:n\geq1
\]
\[
H_{0}e{}_{m,0}=-\lambda(e_{m+1,0}+e_{m-1,0}+e_{m,1}),\:m\geq1
\]
\[
H_{0}e{}_{00}=-\lambda(e_{1,0}+e_{0,1})
\]
\[
Ve_{mn}=\mu\delta(m)\delta(n)e_{m,n},
\]
where $\lambda>0$, $\mu$ are real parameters. That is the particles
interact only if all of them are at the point $0$. In other terms:
\[
(Hf)_{m,n}=-\lambda(f_{m+1,n}+f_{m-1,n}+f_{m,n+1}+f_{m,n-1}),\:n,m\geq1,
\]
\[
(Hf)_{0,n}=-\lambda(f_{1,n}+f_{0,n+1}+f_{0,n-1}),\:n\geq1
\]
\[
(Hf)_{m,0}=-\lambda(f_{m+1,0}+f_{m-1,0}+f_{m,1}),\:m\geq1
\]
\[
(Hf)_{00}=-\lambda(f_{10}+f_{01})+\mu f_{00}
\]

For any selfadjoint operator $A$ we denote $\sigma(A)$, $\sigma_{ess}(A),$
$\sigma_{p}(A),$ $\sigma_{d}(A)$ correspondingly the spectrum, essential
spectrum, point spectrum and discrete spectrum of $A$ (see definitions
in Appendix).

It is well known (Weyl theorem, see \cite{Reed_Simon_4}) that the
essential spectrum of $H$ coincides with the essential spectrum of
$H_{0}$. The following result completely describes the spectrum of
$H_{0}$.

\begin{theorem}\label{L_1} 
\[
\sigma(H_{0})=\sigma_{ess}(H_{0})=[-4\lambda,4\lambda],\sigma_{p}(H_{0})=\text{\ensuremath{\emptyset}}
\]
\end{theorem}

Our main result is the following theorem.

\begin{theorem}\label{th1}
  $\phantom{a}$
  
\begin{itemize}
\item If $\mu\neq0$ and $\left|\frac{\lambda}{\mu}\right|<2\left(1-\frac{8}{3\pi}\right)$
then the discrete spectrum $\sigma_{d}(H)$ coincides with the point
spectrum $\sigma_{p}(H)$ and consists of the unique eigenvalue $E\in(-\infty,-4\lambda)\cup(4\lambda,\infty),$
moreover its sign coincides with the sign of $\mu$. 
\item If either $\mu=0$ or $\left|\frac{\lambda}{\mu}\right|>2\left(1-\frac{8}{3\pi}\right),$
then $\sigma_{d}(H)=\sigma_{p}(H)=\emptyset$. 
\item If $\left|\frac{\lambda}{\mu}\right|=2\left(1-\frac{8}{3\pi}\right)$
then the point spectrum $\sigma_{p}(H)$ consists of the unique eigenvalue,
moreover $E=-4\lambda$ for $\frac{\lambda}{\mu}=-2\left(1-\frac{8}{3\pi}\right)$
and $E=4\lambda$ for $\frac{\lambda}{\mu}=2\left(1-\frac{8}{3\pi}\right);$
$\sigma_{d}(H)=\emptyset.$ 
\end{itemize}
\end{theorem}

\section{Proofs}

\subsection{Essential spectrum: proof of Theorem \ref{L_1}}

\paragraph{Graph spectrum}

Consider countable graph $G=(V,E)$ with the set of vertices $V$
and the set of edges $E.$ Define the following operator (laplacian)
on the space $l_{2}(V):$ 
\begin{equation}
\left(Af\right)(v)=-\lambda\sum_{u\in O_{v}}f(u)\label{laplac}
\end{equation}
where $O_{v}$ is the set of vertices adjacent (neighboring) to vertex
$v.$

Then 
\begin{equation}
\sigma(A)\subseteq[-\lambda\,deg\,G,\lambda\,deg\,G],\label{estimate}
\end{equation}
where $deg\,G$ is the maximal degree of vertices in $G$. It follows
from the obvious upper bound $\lambda deg\,G$ for the norm of $A$.

It follows that $\sigma(H_{0})\subseteq[-4\lambda,4\lambda]$.

\paragraph*{Cartesian product of graphs}

Consider two simple graphs (without cycles and multipl\d{e} edges)
$G_{1},G_{2}.$ and their cartesian product $G_{1}\times G_{2}$.
Let $A$ and $A_{i}$ be laplacians, as in \ref{laplac}, for graphs
$G_{1}\times G_{2}$ and $G_{i}$ correspondingly. Then 
\[
A=A_{1}\times I+I\times A_{2}
\]
The following theorem, proven in \cite{MW}, completely defines the
spectrum of $A$ if we know the spectra of $A_{i}$.

\begin{theorem}\label{mohar} 
\[
\sigma(A)=\{\nu_{1}+\nu_{2}:\nu_{i}\in\sigma(A_{i})\}
\]
\[
\sigma_{p}(A)=\{\nu_{1}+\nu_{2}:\nu_{i}\in\sigma_{p}(A_{i})\}
\]
\end{theorem}

\paragraph*{Spectrum of one-dimensional operator}

In our case we put $G_{i}={\bf Z}_{+}$ $i=1,2$, and assume that
the set $B$ consists of two points $(1,0)$ and $(0,1).$ Also the
operators $A_{i}$ $i=1,2$ are defined to be 
\begin{align*}
\left(A_{i}f\right)(n) & =-\lambda\left(f(n+1)-f(n-1)\right),\:n\geq1\\
\left(A_{i}f\right)(n) & =-\lambda f(1)
\end{align*}
Then
\begin{lemma}
\begin{align*}
\sigma(A_{i}) & =[-2\lambda,2\lambda],\:\sigma_{p}(A_{i})=\emptyset
\end{align*}
\end{lemma}
{\bf Proof} Consider the resolvent equation 
\[
A_{i}f-\nu f=g,
\]
with $f,g\in l_{2}({\bf Z}_{+}).$ If $\nu$ belongs to the resolvent
set, then for any $g\in l_{2}({\bf Z}_{+})$ this equation has unique
solution $f\in l_{2}({\bf Z}_{+}).$ The spectrum is the complement
of the resolvent set. In terms of the generating functions 
\[
F(z)=\sum_{n=0}^{\infty}\,f_{n}z^{n},\;G(z)=\sum_{n=0}^{\infty}\,g_{n}z^{n},
\]
where $f=\{f_{n}\}$ and $g=\{g_{n}\}$, we rewrite the resolvent
equation. After simple transformations we get the equation 
\begin{equation}
F(z)=\frac{\left(z+\frac{\nu}{\lambda}\right)f_{0}+\lambda^{-1}G(z)}{z^{2}+\frac{\nu}{\lambda}z+1}\label{func}
\end{equation}
Using inequality $(\ref{estimate})$ we get that $\sigma(A_{i})$
belongs to $[-2\lambda,2\lambda].$

Let now $\left|\frac{\nu}{\lambda}\right|<2$. We want to show that
$\nu$ does not belong to the resolvent set. For this it is sufficient
to show that the function $\{f_{n}\}$ with the generating function
$(\ref{func})$ does not belong to $l_{2}({\bf Z}_{+}).$ In fact,
quadratic equation in the denominator of (\ref{func}) has two complex
roots $z_{i}$ such that $|z_{i}|=1$. Then we can write 
\[
  F(z)=\frac{(z+\nu /\lambda )f_{0}+\lambda^{-1}G(z)}{(z-z_{1})(z-\bar{z}_{1})}=\frac{(z+\nu / \lambda )f_{0}+
    \lambda^{-1}G(z)}{z_{1}-\bar{z}_{1}}\Bigl(\frac{1}{z-z_{1}}-\frac{1}{z-\bar{z}_{1}}\Bigr)
\]
Then for $|z|<1$ the series 
\begin{align*}
\frac{1}{z-z_{1}} & =-\frac{1}{z_{1} (1- z / z_{1})}=-\frac{1}{z_{1}}\sum_{n=0}^{\infty}\Bigl(\frac{z}{z_{1}}\Bigr)^{n}\\
-\frac{1}{z-\bar{z}_{1}} & =\frac{1}{\bar{z}_{1} (1- z / \bar{z}_{1})}=\frac{1}{\bar{z}_{1}}\sum_{n=0}^{\infty}\left(\frac{z}{\bar{z}_{1}}\right)^{n}
\end{align*}
show that $\nu\in(-2\lambda,2\lambda)$ belongs to the spectrum.

If $\nu=\pm2\lambda,$ then 
\[
F(z)=\frac{\left(z+\frac{\nu}{\lambda}\right)f_{0}+\lambda^{-1}G(z)}{(z\pm1)^{2}}
\]
and one shows similarly that $\{f_{n}\}$ does not belong to $l_{2}({\bf Z}_{+}).$

Similarly one can prove that the discrete spectrum is empty. Then
using Theorem \ref{mohar} we get that $\sigma(H_{0})=[-4\lambda,4\lambda]$
and $\sigma_{p}(H_{0})=\emptyset.$

\subsection{Proof of Theorem \ref{th1}}

The plan of the proof is the following:

1) instead of infinite system of equations we consider equivalent
(functional) equation for generating functions in $\mathbf{C}^{2}$;

2) we project this functional equation to some algebraic curve in
$\mathbf{C}^{2}$, that gives functional equation for two functions
of a real variable;

3) using some transformations and analytic continuation we reduce
this functional equation to the special boundary problem -- the Carleman\tire Dirichlet
problem on the unit circles.

\paragraph*{Functional equation}

Introduce the generating functions 
\begin{align*}
F(x,y) & =\sum_{p,q=1}^{\infty}f_{p,q}x^{p-1}y^{q-1}\\
F(x) & =\sum_{p=1}^{\infty}f_{p,0}x^{p-1}\\
G(y) & =\sum_{q=1}^{\infty}f_{0,q}y^{q-1}
\end{align*}
where $x,y\in\mathbf{C}$. These functions are analytic for $|x|,|y|<1$.
Introduce more convenient (renormalized) parameters 
\[
\nu=\lambda^{-1}E-4,\:\alpha=\lambda{}^{-1}\mu 
\]
Put 
\[
Q(x,y)=y^{2}x+x^{2}y+\nu xy+x+y
\]
\[
q_{1}(x,y)=x^{2}+xy+\nu x+1
\]
\[
q_{2}(x,y)=y^{2}+xy+\nu y+1
\]
\[
q_{0}(x,y)=x+y+\nu-\alpha
\]
\begin{lemma}\label{L_2}
For $|x|<1,|y|<1,$ the following functional equation holds: 
\begin{equation}
-Q(x,y)F(x,y)=q_{1}(x,y)F(x)+q_{2}(x,y)G(y)+q_{0}(x,y)f_{0,0},\label{main_equation}
\end{equation}
\end{lemma}
{\bf Proof} If $f$ is the eigenfunction with eigenvalue $E$ then 
\[
(Hf)_{m,n}=-\lambda(f_{m+1,n}+f_{m-1,n}+f_{m,n+1}+f_{m,n-1})=Ef{}_{m,n},\:n,m\geq1
\]
\[
(Hf)_{0,n}=-\lambda(f_{1,n}+f_{0,n+1}+f_{0,n-1})=Ef_{0,n},\:n\geq1
\]
\[
(Hf)_{m,0}=-\lambda(f_{m+1,0}+f_{m-1,0}+f_{m,1})=Ef_{m,0},\:m\geq1
\]
\[
(Hf)_{0,0}=-\lambda(f_{1,0}+f_{0,1})+\mu f_{0,0}=Ef_{0,0}
\]
We multiply each equation on $\lambda^{-1}x^{m}y^{n}$ and sum up
\[
\left(x+x^{-1}+y+y^{-1}\right)\sum_{m,n=1}^{\infty}f_{m,n}x^{m}y^{n}+\sum_{m=1}^{\infty}f_{m,1}x^{m}+\sum_{n=1}^{\infty}f_{1,n}y^{n}-
\]
\[
-xy\left(\sum_{n=1}^{\infty}f_{0,n}y^{n-1}+\sum_{m=1}^{\infty}f_{m,0}x^{m-1}\right)=\lambda^{-1}E\sum_{m,n=1}^{\infty}f_{m,n}x^{m}y^{n}-\sum_{n=1}^{\infty}f_{1,n}y^{n}-
\]
\[
-\left(y+y^{-1}\right)\sum_{n=1}^{\infty}f_{0,n}y^{n}+f_{0,1}-yf_{0,0}=\lambda^{-1}E\sum_{n=1}^{\infty}f_{0,n}y^{n}-
\]
\[
-\sum_{m=1}^{\infty}f_{m,1}x^{m}-\left(x+x^{-1}\right)\sum_{m=1}^{\infty}f_{m,0}x^{m}+f_{1,0}-xf_{0,0}=\lambda^{-1}E\sum_{m=1}^{\infty}f_{m,0}x^{m}
\]
\[
f_{1,0}+f_{0,1}=-\left(\frac{E-\mu}{\lambda}\right)f_{0,0}
\]
Or
\[
\left(y^{2}x+x^{2}y+\nu xy+x+y\right)F(x,y)+\left(y^{2}+1+xy+\nu y\right)G(y)+
\]
\[
+\left(x^{2}+1+xy+\nu x\right)F(x)+\left(x+y+\nu-\alpha\right)f_{0,0}=0
\]

\paragraph*{Projection onto algebraic curve}

As $q_{1}(y,x)=q_{2}(x,y)\text{,}$ it is sufficient to use one function
\begin{equation}
q(x,y)=x^{2}+xy+\nu x+1\label{qxy}
\end{equation}
Then the main equation is 
\[
-Q(x,y)F(x,y)=q(x,y)F(x)+q(y,x)G(y)+q_{0}(x,y)f_{0,0}
\]

On the algebraic curve $Q(x,y)=0$ the equation is 
\[
q(x,y)F(x)+q(y,x)G(y)+q_{0}(x,y)f_{0,0}=0
\]
or 
\[
-\frac{xF(x)}{y}-\frac{yG(y)}{x}+(x+y+\nu-\alpha)f_{0,0}=0
\]
as 
\[
q(x,y)=-\frac{x}{y},\,\mod Q(x,y),\quad q(y,x)=-\frac{y}{x},\,\mod Q(x,y)
\]
Multiplying on $xy$ and changing the sign we get 
\[
x^{2}F(x)+y^{2}G(y)-(y^{2}x+x^{2}y+(\nu-\alpha)xy)f_{0,0}=0
\]
Taking into account also that 
\[
-(y^{2}x+x^{2}y+(\nu-\alpha)xy)f_{0,0}=-(Q-x-y-\alpha xy)f_{0,0}=(-Q+x+y+\alpha xy)f_{0,0}
\]
we get the equation on the curve $Q(x,y)=0$ 
\[
x^{2}F(x)+y^{2}G(y)+(x+y+\alpha xy)f_{0,0}=0
\]
Put $F_{1}(x)=x\left(f_{0,0}+xF(x)\right),$ $G_{1}(y)=y\left(f_{0,0}+yG(y)\right).$
Then the final equation is 
\begin{equation}
F_{1}(x)+G_{1}(y)+\alpha xyf_{0,0}=0\label{al_cur}
\end{equation}

\paragraph{Case $\boldsymbol{|\nu|>4}$}

Consider equation (\ref{al_cur}) on the part of the curve where $|x|,|y|<1$.
Denote $\Gamma_{x}$($\Gamma_{y}$) the unit circle in the complex
plane $\mathbf{C}_{x}$($\mathbf{C}_{y}$). Denote $x_{1}(y)$ (correspondingly
$y_{1}(x)$) the branch of function $x(y)$ (correspondingly $y(x)$),
satisfying condition $|x_{1}(1)|<1$ ($|y_{1}(1)|<1$).

Consider the equation $Q(x,y)=0.$ It defines two algebraic functions,
$x(y)$ and $y(x)$.

\begin{lemma}\label{L_3}
Both have 4 real branching points.
\end{lemma}
{\bf Proof}
They can be found from the equation $D(x)=0$, where $D(x)=(x^{2}+\nu x+1)^{2}-4x^{2}$
is the discriminant of the quadratic equation $xy^{2}+(x^{2}+\nu x+1)y+x=0.$
As 
\[
D(x)=\left(x^{2}+(\nu+2)x+1\right)\left(x^{2}+(\nu-2)x+1\right)
\]
the branching points are the roots of two quadratic equations
\[
x^{2}+(\nu+2)x+1=0,\,x^{2}+(\nu-2)x+1=0
\]
For $|\nu|>4$ these equations have real roots. Otherwise speaking,
we get 4 branching points. For both equations the product of its roots
is equal to 1. Thus two branching points are inside the unit circle
and other two are outside. We can order these points as $x_{1}<x_{2}<x_{3}<x_{4}.$
Then for $\nu>4$ we have $x_{1}<x_{2}<-1<x_{3}<x_{4}<0$, and for
$\nu<-4$ will be $0<x_{1}<x_{2}<1<x_{3}<x_{4}.$

Consider now the plane with cuts $\widetilde{\mathbf{C}}_{x}=\mathbf{C}_{x}\setminus\left([x_{1},x_{2}]\cup[x_{3},x_{4}]\right).$

\begin{lemma}\label{L_5}Let $|\nu|>4$. Then the function $y(x)$
has two branches $y_{1}(x),y_{2}(x)$ on $\widetilde{\mathbf{C}}_{x}$
such that for all $x\in\widetilde{\mathbf{C}}_{x}$ will be $|y_{1}(x)|<1,$$|y_{2}(x)|>1.$

Moreover, for $|x|=1$ the functions $y_{1}(x),y_{2}(x)$ have real
values and 
\begin{align*}
y_{1}(x) & =\frac{-\left(x+x^{-1}+\nu\right)-\sqrt{\left(x+x^{-1}+\nu-2\right)\left(x+x^{-1}+\nu+2\right)}}{2},\:\nu<-4\\
y_{1}(x) & =\frac{-\left(x+x^{-1}+\nu\right)+\sqrt{\left(x+x^{-1}+\nu-2\right)\left(x+x^{-1}+\nu+2\right)}}{2},\:\nu>4
\end{align*}
Similar formula holds for the function $x_{1}(y).$
\end{lemma}

This lemma follows from Theorem 5.3.3 \cite{FYM} p.\thinspace 137.

By this lemma $x_{1}(y)$ is analytic and $|x_{1}(y)|<1$ in some
neighborhood of $\Gamma_{y}$. Similarly, $y_{1}(x)$ is analytic
and $|y_{1}(x)|<1$ in some neighborhood of $\Gamma_{x}$.

It follows that unknown functions $F_{1}(x)$,$G_{1}(y)$ can be analytically
continued to some neighborhood of the unit circle, and the equation
(\ref{al_cur}) also holds in this neighborhood. Then of course $F_{1}$
and $G_{1}$ are continuous on the unit circle.

That is why for $y\in\Gamma_{y}$ we have 
\begin{align*}
F_{1}(x_{1}(y))+G_{1}(y)+\alpha x_{1}(y)yf_{0,0} & =0\\
F_{1}(x_{1}(y^{-1}))+G_{1}(y^{-1})+\alpha x_{1}(y^{-1})y^{-1}f_{0,0} & =0
\end{align*}
By lemma \ref{L_5} $x_{1}(y)=x_{1}(y^{-1}).$ Subtracting the second
equation from the first one, we get the following equation, that holds
on the unit circle:
\[
G_{1}(y)-G_{1}(y^{-1})+\alpha x_{1}(y)(y-y^{-1})f_{0,0}=0,\:y\in\Gamma_{y}
\]

It will be convenient to use symbol $t$ instead of $y$ if $y\in\Gamma_{y}$:
\begin{equation}
G_{1}(t)-G_{1}\Bigl(\frac{1}{t}\Bigr)+\alpha x_{1}(t)\Bigl(t-\frac{1}{t}\Bigr)f_{0,0}=0,\:t\in\Gamma_{y}\label{b_eq}
\end{equation}

\paragraph{Solution}

Our next problem is to find function $G_{1}(y)$ analytic in the interior
$\mathcal{D}$ of the unit circle and continuous on the unit circle
$\Gamma_{y}$ and such that on $\Gamma_{y}$ the equation (\ref{b_eq})
holds. According to the general theory (see Theorem \ref{T_CD} from
Appendix ) the solution is 
\begin{multline}
G_{1}(y)=\frac{1}{2\pi i}\intop_{\Gamma_{y}}\frac{\alpha x_{1}(\frac{1}{t})\left(\frac{1}{t}-t\right)f_{0,0}}{t-y}dt+C=\\
=-\frac{1}{2\pi i}\intop_{\Gamma_{y}}\frac{\alpha x_{1}(t)\left(t-\frac{1}{t}\right)f_{0,0}}{t-y}dt+C,\;y\in\mathring{\mathcal{D}}\label{d}
\end{multline}
where $\mathring{\mathcal{D}}$ is the interior of the unit circle.

From the condition $G_{1}(0)=0$ we can find constant $C\text{:}$
\[
C=\frac{1}{2\pi i}\intop_{\Gamma_{y}}\frac{\alpha x_{1}(t)\left(t-\frac{1}{t}\right)f_{0,0}}{t}dt
\]
Whence, 
\begin{multline*}
G_{1}(y)=-\frac{1}{2\pi i}\intop_{\Gamma_{y}}\frac{\alpha x_{1}(t)\left(t-\frac{1}{t}\right)f_{0,0}}{t-y}dt+\frac{1}{2\pi i}\intop_{\Gamma_{y}}\frac{\alpha x_{1}(t)\left(t-\frac{1}{t}\right)f_{0,0}}{t}dt=\\
=\frac{1}{2\pi i}\intop_{\Gamma_{y}}\alpha x_{1}(t)\left(t-\frac{1}{t}\right)\left(-\frac{1}{t-y}+\frac{1}{t}\right)f_{0,0}dt=\\
=-\frac{1}{2\pi i}\intop_{\Gamma_{y}}\frac{\alpha yx_{1}(t)\left(t-\frac{1}{t}\right)f_{0,0}}{t\left(t-y\right)}d
\end{multline*}
Thus, 
\[
G_{1}(y)=-\frac{1}{2\pi i}\intop_{\Gamma_{y}}\frac{\alpha yx_{1}(t)\left(t-\frac{1}{t}\right)f_{0,0}}{t\left(t-y\right)}dt,\;y\in\mathring{\mathcal{D}}\Longleftrightarrow
\]
\[
y\left(f_{0,0}+yG(y)\right)==-\frac{1}{2\pi i}\intop_{\Gamma_{y}}\frac{\alpha yx_{1}(t)\left(t-\frac{1}{t}\right)f_{0,0}}{t\left(t-y\right)}dt
\]
It follows that 
\[
f_{0,0}+yG(y)=-\frac{1}{2\pi i}\intop_{\Gamma_{y}}\frac{\alpha x_{1}(t)\left(t-\frac{1}{t}\right)f_{0,0}}{t\left(t-y\right)}dt,\;y\in\mathring{\mathcal{D}}
\]
This equation is equivalent to the following equation 
\[
yG(y)=-\frac{1}{2\pi i}\intop_{\Gamma_{y}}\frac{\alpha x_{1}(t)\left(t-\frac{1}{t}\right)f_{0,0}}{t\left(t-y\right)}dt-f_{0,0}=
\]
\[
=-\frac{1}{2\pi i}\intop_{\Gamma_{y}}\frac{\left(\alpha x_{1}(t)+1\right)\left(t-\frac{1}{t}\right)f_{0,0}}{t\left(t-y\right)}dt
\]
Let us substitute $y=0.$ After canceling $f_{0,0}$ we get the equation
for the eigenvalue $\nu:$ 
\[
-\frac{1}{2\pi i}\intop_{\Gamma_{y}}\frac{x_{1}(t)\left(t-\frac{1}{t}\right)}{t^{2}}dt=\frac{1}{\alpha}
\]
as 
\[
\frac{1}{2\pi i}\intop_{\Gamma_{y}}\frac{t-\frac{1}{t}}{t^{2}}dt=\frac{1}{\alpha}
\]

Finally, we get the following equation for the eigenvalue $\nu$ 
\begin{equation}
-\frac{1}{2\pi i}\intop_{\Gamma_{y}}\frac{x_{1}(t)\left(t-\frac{1}{t}\right)}{t^{2}}dt=\frac{1}{\alpha}=\frac{\lambda}{\mu}\label{nu}
\end{equation}
By lemma \ref{L_5} for $|t|=1$ 
\begin{align*}
x_{1}(t) & =\frac{-\left(t+\nu+t^{-1}\right)-\sqrt{\left(t+t^{-1}+\nu-2\right)\left(t+t^{-1}+\nu+2\right)}}{2},\:\nu<-4\\
x_{1}(t) & =\frac{-\left(t+\nu+t^{-1}\right)+\sqrt{\left(t+t^{-1}+\nu-2\right)\left(t+t^{-1}+\nu+2\right)}}{2},\:\nu>4
\end{align*}
Put 
\[
d(\phi,\nu)=x_{1}(e^{i\phi})=\begin{cases}
-\left(\cos\phi+\frac{\nu}{2}\right)-\sqrt{\left(\cos\phi+\frac{\nu}{2}\right)^{2}-1} & \nu<-4\\
-\left(\cos\phi+\frac{\nu}{2}\right)+\sqrt{\left(\cos\phi+\frac{\nu}{2}\right)^{2}-1} & \nu>4
\end{cases}
\]
and use the following variable change $t=e^{i\phi}$: 
\begin{multline*}
-\frac{1}{2\pi i}\intop_{\Gamma_{y}}\frac{x_{1}(t)\left(t-\frac{1}{t}\right)}{t^{2}}dt=-\frac{1}{2\pi i}\intop_{-\pi}^{\pi}x_{1}(e^{i\phi})2i\sin\phi ie^{-i\phi}d\phi\\
=-\frac{1}{\pi}\intop_{-\pi}^{\pi}d(\phi,\nu)\sin\phi ie^{-i\phi}d\phi=\\
=-\frac{1}{\pi}\intop_{-\pi}^{\pi}d(\phi,\nu)\sin\phi i\cos\phi \, d\phi-\frac{1}{\pi}\intop_{-\pi}^{\pi}d(\phi,\nu)\sin^{2}\phi \, d\phi
\end{multline*}
As the integrand is an odd function, 
\[
\frac{1}{\pi}\intop_{-\pi}^{\pi}d(\phi,\nu)\sin\phi\cos\phi \, d\phi=0
\]
Then 
\[
-\frac{1}{2\pi i}\intop_{\Gamma_{y}}\frac{x_{1}(t)\left(t-\frac{1}{t}\right)}{t^{2}}dt=-\frac{1}{\pi}\intop_{-\pi}^{\pi}d(\phi,\nu)\sin^{2}\phi \, d\phi
\]
and the equation (\ref{nu}) becomes 
\begin{equation}
\frac{1}{\pi}\intop_{-\pi}^{\pi}d(\phi,\nu)\sin^{2}\phi \, d\phi=-\frac{1}{\alpha}\label{nu0}
\end{equation}

The function 
\[
d\left(\phi,\nu\right)=-\cos\phi-\frac{\nu}{2}-\sqrt{\left(\cos\phi+\frac{\nu}{2}\right)^{2}-1}>0,
\]
for all $\phi\in(-\pi,\pi]$ and $\nu<-4$. Moreover, for any $\phi$
it increases monotonically in $\nu$ for $\nu<-4$ and $d\left(\phi,\nu\right)\to0$
as $\nu\to-\infty.$

Then the integral 
\[
\frac{1}{\pi}\intop_{-\pi}^{\pi}d\left(\phi,\nu\right)\sin^{2}\phi \, d\phi
\]
also increases monotonically in $\nu$ and takes values in the interval
$(0,c)$, where 
\begin{align*}
c&=\frac{1}{\pi}\intop_{-\pi}^{\pi}d\left(\phi,-4\right)\sin^{2}\phi \, d\phi=\\
&=\frac{1}{\pi}\intop_{-\pi}^{\pi}\left(-\cos\phi+2-\sqrt{\left(\cos\phi-2\right)^{2}-1}\right)\sin^{2}\phi \, d\phi>0
\end{align*}
It follows that for $0<- \lambda / \mu <c$ there exists a unique
solution $\nu<-4$ of the equation (\ref{nu0}).

One can find the constant $c$ in an explicit form. We have 
\begin{multline*}
\frac{1}{\pi}\intop_{-\pi}^{\pi}\left(-\cos\phi+2-\sqrt{\left(\cos\phi-2\right)^{2}-1}\right)\sin^{2}\phi \, d\phi=\\
=2-\frac{2}{\pi}\intop_{0}^{\pi}\left(\sqrt{\left(\cos\phi-2\right)^{2}-1}\right)\sin^{2}\phi\,d\phi
\end{multline*}
and 
\[
\frac{2}{\pi}\intop_{-\pi}^{\pi}\sin^{2}\phi d\phi=2,\;\frac{1}{\pi}\intop_{-\pi}^{\pi}\cos\phi\sin^{2}\phi \, d\phi=0
\]
Then 
\begin{align*}
&\intop_{0}^{\pi}\left(\sqrt{\left(\cos\phi-2\right)^{2}-1}\right)\,\sin^{2}\phi\,d\phi=\\
&\qquad =\intop_{0}^{\pi}\sqrt{\left(1-\cos\phi\right)\left(3-\cos\phi\right)}\,\sqrt{1-\cos^{2}\phi}\,\sin\phi\,d\phi=\\
&\qquad =\intop_{0}^{\pi}\left(1-\cos\phi\right)\sqrt{\left(3-\cos\phi\right)\left(1+\cos\phi\right)}\sin\phi\,d\phi=\\
&\qquad =\intop_{-1}^{1}\left(1-t\right)\sqrt{\left(3-t\right)\left(1+t\right)}\,dt=\intop_{-1}^{1}\left(1-t\right)\sqrt{4-1+2t-t^{2}}\,dt=\\
&\qquad =\intop_{0}^{2}t\sqrt{4-t^{2}}\,dt=-\frac{1}{3}\left(4-t^{2}\right)^{3/2}\Big|_{0}^{2}=\frac{8}{3}
\end{align*}
Thus 
\begin{equation}
c=2-\frac{16}{3\pi}=2\Bigl(1-\frac{8}{3\pi}\Bigr)\label{cc}
\end{equation}

In order to consider the case $\nu>4$ note that $d\left(\phi,\nu\right)=d\left(-\phi,\nu\right)$
and $d\left(\phi,\nu\right)=-d\left(\pi-\phi,-\nu\right).$ Then we
have 
\[
\frac{1}{\pi}\intop_{-\pi}^{\pi}d\left(\phi,\nu\right)\sin^{2}\phi d\phi=\frac{2}{\pi}\intop_{0}^{\pi}d\left(\phi,\nu\right)\sin^{2}\phi \, d\phi
\]
and 
\[
\frac{2}{\pi}\intop_{0}^{\pi}d\left(\phi,\nu\right)\sin^{2}\phi d\phi=-\frac{2}{\pi}\intop_{0}^{\pi}d\left(\phi,-\nu\right)\sin^{2}\phi \, d\phi
\]
Thus, the integral 
\[
\frac{1}{\pi}\intop_{-\pi}^{\pi}d\left(\phi,\nu\right)\sin^{2}\phi \, d\phi,
\]
is an odd function of $\nu$ and, hence, it is monotonically increasing
for $\nu>4$ and takes its values in $(-c,0).$ So for $-c<-\frac{\lambda}{\mu}<0,$
the equation (\ref{nu0}) has a unique solution $\nu>4.$

Thus, for $\left|\frac{\lambda}{\mu}\right|<c=2\left(1-\frac{8}{3\pi}\right)$
the equation (\ref{nu0}) has unique solution $\nu$, $|\nu|>4,$
where the sign of $\nu$ coincides with the sign of $\mu.$

\paragraph{Case $\boldsymbol{|\nu|\leq4}$}

Here we consider the case $|\nu|\leq4$ and prove the last assertion
of the Theorem.

\paragraph{Preliminary lemmas}

We will need some properties of the algebraic functions $x(y)$ and
$y(x),$ defined by the equation $Q(x,y)=0$ for $|\nu|\leq4.$

\begin{lemma}\label{L7}
For $|\nu|<4,$ $\nu\neq0,$ the algebraic functions $x(y)$ and $y(x)$
have each 4 branching points: 2 real (inside and outside the unit
circle) and 2 complex conjugate on the unit circle.
\end{lemma}

{\bf Proof} The branching points of $x(y)$ can be found from the equation
$D(y)=0,$ where $D(y)=(y^{2}+\nu y+1)^{2}-4y^{2}$ is the discriminant
of the quadratic equation $yx^{2}+(y^{2}+\nu y+1)x+y=0.$ As 
\[
D(y)=\left(y^{2}+(\nu+2)y+1\right)\left(y^{2}+(\nu-2)y+1\right)
\]
we have two pairs of roots easily found.

Denote by $y_{1},y_{4}$ real roots, and by $y_{2},y_{3}$ the complex ones, so that
$y_{2}=\bar{y}_{3},$ $|y_{2}|=|y_{3}|=1.$ Assume that $|y_{1}|<|y_{4}|$
and $Im\,y_{2}>0.$ For $-4<\nu<0$ we have $0<y_{1}<1<y_{4}$, and
for $0<\nu<4$ we have $y_{4}<-1<y_{1}<0\text{.}$

Firstly, consider the case $|\nu|=4.$

\begin{lemma}\label{L8}
For $\nu=-4$ we have $y_{2}=y_{3}=1,$ $0<y_{1}<1<y_{4},$ and for
$\nu=4$ we have $y_{2}=y_{3}=-1,$ $y_{1}<-1<y_{4}<0.$
\end{lemma}

\begin{lemma}\label{L9}
For $|\nu|\leq4$ the algebraic function $x(y)$ has two branches
$x_{1}(y),$ $x_{2}(y)$ such that for $|y|=1$ $|x_{1}(y)|\leq1$
and $|x_{2}(y)|\geq1.$

Similar statement holds for the algebraic function $y(x).$
\end{lemma}

\paragraph{Operator $\boldsymbol{H_{0}}$}

Here we suggest new method, different from above, to prove that $H_{0}$
does not have point spectrum. It will follow from the following

\begin{proposition}
\label{pp}
  The equation 
\begin{equation}
F_{1}(x)+G_{1}(y)=0,\: |x|<1,\,|y|<1,\quad F_{1}(0)=0,\,G_{1}(0)=0\label{eq-1}
\end{equation}
on the algebraic curve $Q(x,y)=0$ does not have nonzero solutions
for $|\nu|\leq4,$ $\nu\neq0$ in the class of functions which have
power series such that the vector of their coefficients belongs to
$l_{2}$.
\end{proposition}

{\bf Proof} Let for example $-4\leq\nu<0.$

By Lemmas \ref{L7} and \ref{L8} there exist branching points $y_{1},$
$y_{2}$, such that $0<y_{1}\leq1$ and $|y_{2}|=1,$ $Im\,y_{2}\geq0.$
For $\nu=-4$ we have $0<y_{1}<1,$$y_{2}=1$ and for $-4<\nu<0$
will be $0<y_{1}<1,$ $Im\,y_{2}>0.$

For $-4<\nu<0$ let us consider the cut (incision) $\gamma$ starting
from the point $y_{1}$ to point $1$ along the real axis and continue
it along the circle from point $1$ until point $y_{2}.$ For $\nu=-4$
put $\gamma=[y_{1},1].$

Note that the function $x_{1}(y)$ maps $\gamma$ onto the unit circle.
Note also that $x_{1}(y_{1})=-1,$ $x_{1}(y_{2})=1,$ $x_{1}(1)=y_{2}.$
Moreover, $|x_{1}(y)|<1$ for $|y|<1,$ if $y\notin\gamma.$

Substituting now $x=x_{1}(y)$ and $y=y_{1}(x)$ to the equation
(\ref{eq-1}) we get 
\begin{align}
F_{1}(x_{1}(y))+G_{1}(y) & =0\label{e1}\\
F_{1}(x)+G_{1}(y_{1}(x)) & =0\nonumber 
\end{align}
where $|x|<1,$ $|y|<1$ and the branches $x_{1}(y),y_{1}(x)$ were
defined in lemma \ref{L9}.

By symmetry $Q(x,y)=Q(y,x)$ we have equality $y_{1}(y)=x_{1}(y).$
From the second equation in \ref{e1} we have 
\begin{equation}
G_{1}(x_{1}(y))+F_{1}(y)=0,\;|y|<1\label{e2}
\end{equation}
Subtracting (\ref{e1}) from (\ref{e2}) we have 
\begin{equation}
h(y)-h(x_{1}(y))=0,\:|y|<1\label{e2-1}
\end{equation}
for 
\[
h(y)=F_{1}(y)-G_{1}(y),\:|y|<1
\]

We will need the following auxiliary assertion.

Let $\mathcal{H}_{2}$ be the Hardy space, see \cite{Duren}. Note
that the function $h(y)\in\mathcal{H}_{2}$ means that $h(y)$ is
analytic function for $|y|<1$ and $\sum_{n=0}^{\infty}|h_{n}|^{2}<\infty.$

\begin{lemma} \label{l10}Let $h(y)\in\mathcal{H}_{2}$ and $h(0)=0$.
Then the equations 
\[
h(y)-\delta h(x_{1}(y))=0,\:h(0)=0,\:\delta=\pm1
\]
do not have nonzero solutions in the domain $|y|<1,\:Im\,y\geq0$.
\end{lemma}

{\bf Proof} Let $g(y)$ be the integral of $h(y)$ so that $h(y)=g^{\prime}(y),$
$g(0)=0.$ Then $g(y)=\sum_{n=2}^{\infty}\,\frac{h_{n-1}}{n}y^{n}$
and the equation will look as follows
\[
g^{\prime}(y)-\delta g^{\prime}(x_{1}(y))=0
\]
Note that the function $g(y)$ is analytic for $|y|<1$ and is continuous
on the unit circle. In fact, the convergence of the series $\sum_{n=1}^{\infty}\,|h_{n}|^{2}$
implies the convergence of the series $\sum_{n=2}^{\infty}\,\frac{|h_{n-1}|}{n}.$

Multiply both sides of this equation on $1-y^{-2}$ and integrate
it along any path from $y_{1}$ to $y,$ not intersecting the cut.
We get 
\[
\int_{y_{1}}^{y}\left(1-t^{-2}\right)dg(t)=\delta\int_{y_{1}}^{y}\left(1-t^{-2}\right)\frac{dg(x_{1}(t))}{x_{1}^{\prime}(t)}
\]
for $|y|<1,\:Im\,y>0.$ As 
\[
\frac{1}{x_{1}^{\prime}(t)}=\frac{x_{1}^{-2}(t)-1}{1-t^{-2}}
\]
we have
\begin{equation}
\int_{y_{1}}^{y}\left(1-t^{-2}\right)dg(t)=\delta\int_{y_{1}}^{y}\left(x_{1}^{-2}(t)-1\right)dg(x_{1}(t))\label{eeq}
\end{equation}
Denote 
\begin{align*}
l(t) & =1-t^{-2}\\
m(t) & =x_{1}^{-2}(t)-1
\end{align*}
Then 
\[
\int_{y_{1}}^{y}l(t)dg(t)=\delta\int_{y_{1}}^{y}m(t)dg(x_{1}(t))
\]
Integrate each of both integrals by part and use that $m(y_{1})=x_{1}^{-2}(y_{1})-1=0:$
\begin{multline}
l(y)g(y)-l(y_{1})g(y_{1})-\int_{y_{1}}^{y}l^{\prime}(t)g(t)dt=\\
=\delta\Biggl(m(y)g(x_{1}(y))-\int_{y_{1}}^{y}m^{\prime}(t)g(x_{1}(t))dt\Biggr)\label{eq}
\end{multline}
where 
\begin{align*}
l^{\prime}(t) & =\frac{2}{t^{3}}\\
m^{\prime}(t) & =-\frac{2}{x_{1}^{3}(t)}\times\frac{1-t^{-2}}{x_{1}^{-2}(t)-1}=-\frac{2\left(1-t^{-2}\right)}{x_{1}(t)\left(1-x_{1}^{2}(t)\right)}
\end{align*}
We see that the equation (\ref{eq}) can be continued to the semicircle
$Im\,y\geq0,$ as the integral 
\[
\int_{y_{1}}^{y}m^{\prime}(t)g(x_{1}(t))dt
\]
exists. In fact, the integrand has singularities at the points $y_{1}$,
$y_{2}.$ For $-4<\nu<0$ the points $y_{1}$, $y_{2}$ are branching
points of order 2, and $x_{1}(y_{1})=-1$, $x_{1}(y_{2})=1.$ Thus,
in a neighborhood of $y_{1}$ we have $x_{1}(y)+1\sim c_{1}(y-y_{1})^{1/2}$,
and in a neighborhood of $y_{2}$ we have $x_{1}(y)-1\sim c_{2}(y-y_{2})^{1/2}$.
It follows that the integral exists. If $\nu=-4$ then $y_{1}$ is
also second order branching point. And in the point $y_{2}=1$ the
function $x_{1}(y)$ behaves as $x_{1}(y)-1\sim c_{2}(y-1)$. Due
to the numerator this is a removable singularity.

Denote by $\Gamma_{1}$ the arc $(y_{2},1)$ of the semicircle $Im\,y\geq0$.
Consider the equation (\ref{eq}) for $y\in\Gamma_{1}.$ Choose the
path from $y_{1}$ to $y$ along the internal side of the cut. Then
\begin{multline}
l(y)g(y)-l(y_{1})g(y_{1})-\int_{y_{1}}^{y}l^{\prime}(t)g(t)dt=\\
= \delta\left(x_{1}^{-2}(y)-1\right)g(x_{1}(y))+\delta\int_{y_{1}}^{y}\frac{2\left(1-t^{-2}\right)g(x_{1}(t))}{x_{1}(t)\left(1-x_{1}^{2}(t)\right)}dt\label{e3}
\end{multline}
Now consider the path which goes initially from $y_{1}$ to $y_{2}$
along the internal side of the cut $\gamma,$ and then along the external
side of this cut from $y_{2}$ to $y$. Then the equation (\ref{eq})
can be written as follows 
\begin{align}
  &l(y)g(y)-l(y_{1})g(y_{1})-\int_{y_{1}}^{y}l^{\prime}(t)g(t)dt= \nonumber \\[4pt]
  &\qquad = \delta\left(x_{2}^{-2}(y)-1\right)g(x_{2}(y))+ \nonumber \\[4pt]
&\qquad\qquad {} +\delta\int_{y_{1}}^{y_{2}}\frac{2\left(1-t^{-2}\right)g(x_{1}(t))}{x_{1}(t)\left(1-x_{1}^{2}(t)\right)}dt+\delta\int_{y_{2}}^{y}\frac{2\left(1-t^{-2}\right)g(x_{2}(t))}{x_{2}(t)\left(1-x_{2}^{2}(t)\right)}dt\label{e4}
\end{align}
Remind that $x_{1}(t)x_{2}(t)=1.$ Then, subtracting equation (\ref{e4})
from the equation (\ref{e3}) we get 
\begin{multline}
\delta\left(\left(x_{1}^{-2}(y)-1\right)g(x_{1}(y))-\left(x_{2}^{-2}(y)-1\right)g(x_{2}(y))\right)=\\
=\delta2\int_{y}^{y_{2}}\left(1-t^{-2}\right)\biggl(\frac{g(x_{1}(t))}{x_{1}(t)\left(1-x_{1}^{2}(t)\right)}-\frac{g(x_{1}^{-1}(t))}{x_{1}^{-1}(t)\left(1-x_{1}^{-2}(t)\right)}\biggr)dt\label{e5}
\end{multline}
Consider the integral in the right hand side of the equation (\ref{e5}).
We use the change of variable $s=x_{1}(t).$ Then $t=y_{1}(s)$ and
\[
dt=\frac{1-s^{-2}}{y_{1}^{-2}(s)-1}\,ds
\]
After this the integral in the right hand side of (\ref{e5}) will
look 
\begin{multline*}
\int_{x_{1}(y)}^{x_{1}(y_{2})}\left(1-y_{1}^{-2}(s)\right)\left(\frac{g(s)}{s\left(1-s^{2}\right)}-\frac{sg(s^{-1})}{1-s^{-2}}\right)\frac{1-s^{-2}}{y_{1}^{-2}(s)-1}\,ds=\\
=\int_{x_{1}(y)}^{1}\,s^{-3}g(s)ds+\int_{x_{1}(y)}^{1}\,sg(s^{-1})ds=\\
=\int_{x_{1}(y)}^{1}\,s^{-3}g(s)ds-\int_{x_{1}(y)}^{1}\,s^{-3}g(s)ds=\\
=\int_{x_{1}(y)}^{x_{2}(y)}\,s^{-3}g(s)ds\qquad\qquad\qquad\qquad\qquad\qquad\qquad\qquad\;\,\,
\end{multline*}
as $x_{2}(y_{2})=1.$ Then the equation (\ref{e5}) (after cancellation
of $\delta$) can be written as 
\[
\left(\left(x_{1}^{-2}(y)-1\right)g(x_{1}(y))-\left(x_{2}^{-2}(y)-1\right)g(x_{2}(y))\right)=\int_{x_{1}(y)}^{x_{2}(y)}\,s^{-3}g(s)ds
\]
Putting $z=x_{1}(y)$ with $y\in\Gamma_{1},$ we get the equation
\begin{equation}
\left(1-z^{-2}\right)g(z)-\left(1-\bar{z}^{-2}\right)g(\bar{z})=2\int_{\bar{z}}^{z}\,s^{-3}g(s)ds\label{e6}
\end{equation}
where $z$ belong to the arc$(y_{2},1,\bar{y}_{2})$ of the unit circle.

Now we want to analytically continue this equation to the whole circle.
Let $y$ belong to the arc $(y_{2},-1,\bar{y}_{2}).$ On this arc
the values of the function $x_{1}(y)$ belong to the cut $(y_{1},1)$,
where the function $h$ is continuous. That is why we have $h(y)=h(\bar{y})$
for $y\in(y_{2},-1,\bar{y}_{2}).$ It follows 
\[
g^{\prime}(y)=g^{\prime}(y^{-1})
\]
Multiply both sides of this equation on $1-y^{-2}$ and integrate
along the arc $(-1,y)$ of the circle: 
\[
\int_{-1}^{y}\left(1-t^{-2}\right)g^{\prime}(t)dt=\int_{-1}^{y}\left(1-t^{-2}\right)g^{\prime}(t^{-1})dt
\]
The integral in the right hand part of this equality is equal to the
integral 
\[
\int_{-1}^{\bar{y}}\left(1-t^{-2}\right)g^{\prime}(t)dt
\]
Then we get the equation 
\[
\int_{-1}^{y}\left(1-t^{-2}\right)g^{\prime}(t)dt=\int_{-1}^{\bar{y}}\left(1-t^{-2}\right)g^{\prime}(t)dt
\]
Integrating by parts we have 
\[
\left(1-y^{-2}\right)g(y)-2\int_{-1}^{y}t^{-3}g(t)dt=\left(1-\bar{y}^{-2}\right)g(\bar{y})-2\int_{-1}^{\bar{y}}t^{-3}g(t)dt
\]
Thus 
\begin{equation}
\left(1-y^{-2}\right)g(y)-\left(1-\bar{y}^{-2}\right)g(\bar{y})=2\int_{-1}^{y}t^{-3}g(t)dt+2\int_{\bar{y}}^{-1}t^{-3}g(t)dt=2\int_{\bar{y}}^{y}t^{-3}g(t)dt\label{e7}
\end{equation}
and the equation (\ref{e6}) holds on the whole circle.

Put 
\[
g(y)=y^{2}(c_{1}+ys(y))
\]
and 
\begin{equation}
S(y)=(y^{2}-1)(c_{1}+ys(y))-2\int_{0}^{y}s(v)dv\label{eqS}
\end{equation}
Then the equation (\ref{e7}) can be written as 
\[
S(y)-S(\bar{y})=2c_{1}\ln\frac{y}{\bar{y}},\quad |y|=1,
\]
where the function $S(y)$ is analytic for $|y|<1$ and continuous
for $|y|=1.$ By theorem \ref{T_CD} the solution of this Dirichlet\tire Carleman
boundary value problem is 
\[
S(y)=c+\frac{2c_{1}}{2\pi i}\int_{\Gamma}\frac{\ln (t / \bar{t})}{t-y}\,dt=c+\frac{2c_{1}}{\pi}\int_{\Gamma}\frac{\arg t}{t-y}\,dt
\]
As for $|x|<1$ the equation 
\[
h(x)-\delta h(y_{1}(x))=0
\]
holds and $|y_{1}(x)|<1$ in some neighborhood of the point $x=-1,$
the function $h(x)$ can be analytically continued to some neighborhood
of the point $-1.$ Thus, the function $g$ is also analytic in some
neighborhood of the point $-1.$ It follows that the function $S(y)$
is differentiable at $y=-1:$ 
\[
S^{\prime}(-1)=\lim_{y\to-1,\,|y|<1}\frac{2c_{1}}{\pi}\int_{\Gamma}\frac{\arg t}{\left(t-y\right)^{2}}\,dt
\]
But as this limit does not exist, it follows that $c_{1}=0$ and due
to (\ref{eqS}) 
\[
S(y)=(y^{2}-1)ys(y)-2\int_{0}^{y}s(v)dv=c
\]
for some constant $c.$ Since $S(0)=0,$ we have $c=0.$ We get the differential
equation 
\[
(y^{2}-1)yu^{\prime}(y)-2u(y)=0
\]
for $u(y)=\int_{0}^{y}s(v)dv.$ Its solution is 
\[
u(y)=C\frac{y^{2}-1}{y^{2}}
\]
There is singularity at $0$, hence $C=0.$ It follows also that
$s(y)=0,$ then $g(y)=0$ and $h(y)=0.$  Lemma \ref{l10} is proved.

Now applying this lemma (with $\delta=1$) to equation (\ref{e2-1}),
we get that $F_{1}(y)=G_{1}(y).$ Then we can apply lemma \ref{l10}
(with $\delta=-1$) to the equation 
\begin{equation}
F_{1}(x)+F_{1}(y)=0,\:|x|<1,\,|y|<1,\:F_{1}(0)=0\label{eq-1-1}
\end{equation}
defined on the curve $Q(x,y)=0$, and deduce from this that $F_{1}(x)\equiv0.$

Thus, we proved the assertion of Proposition \ref{pp} for $-4\leq\nu<0.$ The case $0<\nu\leq4$
is quite similar.

\paragraph{Operator $\boldsymbol{H}$}

Now we consider the point spectrum $\sigma_{p}(H)$ of $H$.

\begin{proposition}
If $\frac{\lambda}{\mu}=2\left(1-\frac{8}{3\pi}\right)$ or $\frac{\lambda}{\mu}=-2\left(1-\frac{8}{3\pi}\right),$
then the point spectrum $\sigma_{p}(H)$ consists of the unique eigenvalue
$E=4\lambda$ or $E=-4\lambda.$
\end{proposition}

\noindent
{\it Proof}. For $|\nu|\leq4$, $\nu\neq0$ consider the main equation 
\begin{equation}
F_{1}(x)+G_{1}(y)+\alpha xyf_{0,0}=0\label{ee}
\end{equation}
on the algebraic curve $Q(x,y)=0,$ where $|x|<1$. Contrary to the
case $|\nu|>4$, already considered, the algebraic functions $x(y)$
and $y(x)$, defined by the equation $Q(x,y)=0$, have branch points
on the unit circle and take values with modulus 1 on some arc of the
unit circle. Thus, the functions $F_{1}(x),$ $G_{1}(y)$ cannot be
analytically continued to a neighborhood of the unit circle.

Nevertheless, by theorem \ref{T_CD} there exists solution $F_{2}(x)$,$G_{2}(y)$
of the main equation (\ref{ee}) in the class of functions analytic
inside the unit circle and continuous on its boundary. This solution
looks like before, namely,
\begin{align}
F_{2}(x) & =\frac{1}{2\pi i}\intop_{\Gamma}\frac{\alpha xy_{1}(t)\left(t-\frac{1}{t}\right)f_{0,0}}{t\left(t-x\right)}dt,\;x\in\mathring{\mathcal{D}}\nonumber \\
G_{2}(y) & =\frac{1}{2\pi i}\intop_{\Gamma}\frac{\alpha yx_{1}(t)\left(t-\frac{1}{t}\right)f_{0,0}}{t\left(t-y\right)}dt,\;y\in\mathring{\mathcal{D}}\label{gf}
\end{align}
Thus, the general solution of the main equation (\ref{ee}) in the
Hardy space $\mathcal{H}_{2}$ can be written as 
\begin{align*}
F_{1}(x) & =F_{0}(x)+F_{2}(x)\\
G_{1}(y) & =G_{0}(y)+G_{2}(y)
\end{align*}
where $F_{0}(x),G_{0}(x)$ is the solution of the homogeneous equation
\[
F_{0}(x)+G_{0}(y)=0
\]
Since the homogeneous equation does not have nonzero solutions, the
unique solution of the main equation (\ref{ee}) will be $F_{2}(x),G_{2}(y).$
Then $E$ is an eigenvalue of the hamiltonian $H$ iff $\nu=\lambda^{-1}E$
satisfies the equation 
\begin{equation}
\frac{1}{2\pi i}\intop_{\Gamma_{y}}\frac{x_{1}(t)\left(t-\frac{1}{t}\right)}{t^{2}}dt=-\frac{1}{\alpha}=-\frac{\lambda}{\mu}\label{e}
\end{equation}
We have proved above (see (\ref{cc})) that for $\nu=\mp4$ the right
hand side of this equation is 
\[
\frac{1}{2\pi i}\intop_{\Gamma_{y}}\frac{x_{1}(t)\left(t-1/t\right)}{t^{2}}dt=\pm2\Bigl(1-\frac{8}{3\pi}\Bigr)
\]
It follows that for $|\alpha|=2\left(1-\frac{8}{3\pi}\right)$ $E=\pm4\lambda$
are the eigenvalues. The corresponding eigenvectors are given by generating
functions (\ref{gf}) and belong to $l_{1}({\bf Z}_{+}^{2}).$

One can show that for $|\nu|<4,$ $\nu\neq0$ the left side of the
equation (\ref{e}) takes complex values. It follows that this equation
does not have solution.

To finish the proof we have to show only that $\nu=0$ is not an eigenvalue.
In this case the polynomial $Q(x,y)$ can be factorized 
\begin{align*}
Q(x,y) & =y^{2}x+x^{2}y+x+y=(x+y)(xy+1)
\end{align*}
If the equation (\ref{main_equation}) has a solution then the function
\[
F(x,y)=\frac{F_{1}(x)+G_{1}(y)+\alpha xyf_{0,0}}{(x+y)(xy+1)}
\]
is analytic for $|x|<1,|y|<1.$

Put $x=-y.$ From analyticity of $F(x,y)$ it follows that $F_{1}(-y)+G_{1}(y)-\alpha y^{2}f_{0,0}=0.$
Without loss of generality we can assume that $f_{0,0}=1.$ As $G_{1}(y)=\alpha y^{2}-F_{1}(-y),$
we have 
\[
F(x,y)=\frac{F_{1}(x)-F_{1}(-y)+\alpha y(x+y)}{(x+y)(xy+1)}
\]
Let 
\[
F_{1}(x)=\sum_{k=1}^{\infty}a_{k}x^{k}
\]
Then, using the expansions 
\begin{align*}
\frac{F_{1}(x)-F_{1}(-y)}{x+y} & =\sum_{k=1}^{\infty}a_{k}\sum_{i=0}^{k-1}x^{k-1-i}(-1)^{i}y^{i}\\
\frac{1}{xy+1} & =\sum_{i=0}^{\infty}(-1)^{i}y^{i}x^{i}
\end{align*}
we find 
\[
F(x,y)=\sum_{m,n=0}^{\infty}A_{m,n}x^{m}y^{n}
\]
where 
\[
A_{m,n}=(-1)^{n}\biggl(\alpha\delta_{m,n-1}+\sum_{k=|m-n|+1}^{m+n+1}a_{k}\biggr)
\]
and the last sum is over $k$ such that $m+n-k$ is odd.

Let $m\geq n,$ $m=n+l.$ Then 
\[
A_{n+l,n}=(-1)^{n}\sum_{r=0}^{n}a_{l+1+2r}
\]
As the series is convergent, 
\[
\sum_{n,l\geq0}^{\infty}\left(A_{n+l,n}\right)^{2}=\sum_{n,l\geq0}^{\infty}\biggl(\sum_{r=0}^{n}a_{l+1+2r}\biggr)^{2}<\infty,
\]
  the following series is also convergent for any $l$:
\[
\sum_{n=0}^{\infty}\biggl(\sum_{r=0}^{n}a_{l+1+2r}\biggr)^{2}<\infty
\]
and, thus, as $n\to\infty$, 
\[
S_{l,n}=\sum_{r=0}^{n}a_{l+1+2r}\to 0
\]
It follows that all $a_{k}=0,$ since 
\[
a_{l+1}=S_{l,n}-S_{l+2,n}.
\]
Theorem \ref{th1} is proved.

\appendix

\section{Appendix }

\subsection{Definitions concerning spectrum }

Here, for the reader's convenience, we recall some definitions from \cite{Reed_Simon_4},
 pp.\thinspace 188, 231, 236. We restrict ourselves here to bounded
selfadjoint operators $H$ in the complex Hilbert space $\mathbf{H}$.
The resolvent set $\rho(H$) of $H$ is the set of all $z\in C$ such
that the mapping $zI-H$ is one-to-one (then, by inverse mapping theorem,
it has bounded inverse). The set $\sigma(H)=C\setminus\rho(H)$ is
called the spectrum of $H$. For selfadjoint operators the spectrum
belongs to the real axis. Point $E\in C$ is an eigenvalue if there
exists $x\in\mathbf{H}$, called eigenvector, such that $Hx=Ex$.
The set $\sigma_{p}(H)$ of all eigenvalues is called the point spectrum
of $H$. Note that if $z$ is a isolated point of the spectrum then
it is an eigenvalue (see Proposition on page 236 of {[}2{]}). The
discrete spectrum is the set $\sigma_{d}(H)\subset\sigma_{p}(H)$
of the eigenvalues $z$ such that $z$ is isolated point of the spectrum
and has finite multiplicity, that is the set of the corresponding
eigenvectors has finite dimension. The set $\sigma_{ess}(H)=\sigma(H)\setminus\sigma_{d}(H)$
is called the essential spectrum.

In this paper we get complete description of discrete, point and essential
spectrum.

\subsection{Carleman\tire Dirichlet problem}

Let $\mathcal{L}$ be simple closed smooth contour (in our case it
is the unit circle). The problem is to find function $F(x)$ analytic
inside the circle $\mathcal{L}$ and such that its limiting values
on the circle are continuous and satisfy the equation 
\[
F(\alpha(t))-F(t)=g(t)
\]
where 
\begin{itemize}
\item the function $g(t)$ satisfies the H\"older condition for $t\in\mathcal{L}$ 
\item $\alpha(t)$ is one-to-one mapping of the contour $\mathcal{L}$ on
itself such that this mapping changes the orientation of this contour,
the derivative $\alpha^{\prime}(t)\neq0$ for all $t\in\mathcal{L}$
and $\alpha^{\prime}(t)$ also satisfies H\"older condition. 
\item $\alpha(\alpha(t))=t$ 
\end{itemize}
In our case $\mathcal{L}=\Gamma,$ where $\Gamma$ is the unit circle
and $\alpha(t)=\frac{1}{t}.$

The following theorem was proved in \cite{FYM}, p.\ 132.

\begin{theorem}\label{T_CD}The Carleman\tire Dirichlet problem 
\[
F(\alpha(t))-F(t)=g(t)
\]
where $g(t)$ satisfies the equation $g(\alpha(t))+g(t)=0$, has unique
solution up to some constant 
\[
F(x)=\frac{1}{2\pi i}\int_{\mathcal{L}}\frac{\phi(\alpha(t))}{t-x}dt+C
\]
where $C$ is a constant and $\phi(t)$ is the unique solution
of the integral equation 
\begin{equation}
(\mathcal{B}\phi)(t)\equiv\phi(t)+\frac{1}{2\pi i}\int_{\mathcal{L}}\Bigl(\frac{1}{s-t}-\frac{\alpha^{\prime}(s)}{\alpha(s)-\alpha(t)}\Bigr)\phi(s)ds=g(t)\label{T5_equ}
\end{equation}
\end{theorem}

If $\alpha(t)=1/t$ and $\mathcal{L}=\Gamma,$ then a general
solution of the Carleman\tire Diriochlet problem is 
\[
F(x)=-\frac{1}{2\pi i}\int_{\Gamma}\frac{g(t)}{t-x}dt+C
\]
Indeed,
\[
\frac{1}{s-t}-\frac{\alpha^{\prime}(s)}{\alpha(s)-\alpha(t)}=\frac{1}{s}
\]
and the equation (\ref{T5_equ}) for $\phi(t)$ will look as follows:
\[
\phi(t)+\frac{1}{2\pi i}\int_{\Gamma}\frac{\phi(s)}{s}ds=g(t)
\]
The solution of this equation is the function $\phi(t)=g(t)+c,$ where
\[
c=-\frac{1}{4\pi i}\int_{\Gamma}\frac{g(s)}{s}ds
\]
In fact 
\begin{align*}
g(t)+c+\frac{1}{2\pi i}\int_{\Gamma}\frac{g(s)+c}{s}ds & =g(t)\Longleftrightarrow\\
c+\frac{1}{2\pi i}\int_{\Gamma}\frac{g(s)+c}{s}ds & =0\Longleftrightarrow\\
c & =-\frac{1}{4\pi i}\int_{\Gamma}\frac{g(s)}{s}ds
\end{align*}
According to the theorem the general solution is 
\[
F(x)=\frac{1}{2\pi i}\int_{\Gamma}\frac{g(\alpha(t))+c}{t-x}dt+C=\frac{1}{2\pi i}\int_{\Gamma}\frac{g(\alpha(t))}{t-x}dt+c+C=
\]
\[
=-\frac{1}{2\pi i}\int_{\Gamma}\frac{g(t)}{t-x}dt+C^{\prime}
\]
as $g(\alpha(t))+g(t)=0.$ 

\end{document}